\documentclass[12pt]{article}

\usepackage{cite}
\usepackage{epsfig}

\input paperdef

\begin{document}
\thispagestyle{empty}

\def\thefootnote{\fnsymbol{footnote}}

\begin{flushright}

DCPT/02/126 \hfill
IPPP/02/63\\
LMU 11/02 \hfill
MPI-PhT/2002-73\\
RM3-TH/02-19 \hfill
LC-TH-2002-015\\
hep-ph/0212020 \\
\end{flushright}

\vspace{1cm}

\begin{center}

{\Large\sc {\bf Towards High-Precision Predictions}}

\vspace{0.4cm}

{\Large\sc {\bf for the MSSM Higgs Sector}}
 
\vspace{1cm}

{\sc 
G.~Degrassi$^{1}$%
\footnote{email: giuseppe.degrassi@roma3.infn.it}%
, S.~Heinemeyer$^{2}$%
\footnote{email: Sven.Heinemeyer@physik.uni-muenchen.de}%
, W.~Hollik$^{3}$%
\footnote{email: hollik@mppmu.mpg.de}%
, P.~Slavich$^{3,4}$%
\footnote{email: slavich@mppmu.mpg.de}%
~and G.~Weiglein$^{5}$%
\footnote{email: Georg.Weiglein@durham.ac.uk}
}

\vspace*{1cm}

{\sl
$^1$ Dipartimento di Fisica, Universit\`a di Roma III and 
INFN, Sezione di Roma III,\\ Via della Vasca Navale~84, I--00146 Rome, Italy 
 
\vspace*{0.4cm}

$^2$Institut f\"ur Theoretische Elementarteilchenphysik,
LMU M\"unchen, \\ Theresienstr.\ 37, D--80333 Munich, Germany

\vspace*{0.4cm}

$^3$Max-Planck-Institut f\"ur Physik (Werner-Heisenberg-Institut),\\
F\"ohringer Ring 6, D--80805 Munich, Germany

\vspace*{0.4cm}

$^4${Institut f\"ur Theoretische Physik, Universit\"at Karlsruhe,\\
Kaiserstrasse 12, Physikhochhaus, D--76128 Karlsruhe, Germany}

\vspace*{0.4cm}

$^5$Institute for Particle Physics Phenomenology, University of Durham,\\
Durham DH1~3LE, UK

}

\end{center}

\vspace*{1cm}

\begin{abstract}
The status of the evaluation of the MSSM Higgs sector is reviewed. The
phenomenological impact of recently obtained corrections is discussed.
In particular it is shown that the upper bound on $\mh$ within the MSSM 
is shifted upwards. Consequently, lower limits on $\Tb$ obtained by
confronting the upper bound as function of $\tb$ with the lower bound on
$\mh$ from Higgs searches are significantly weakened.
Furthermore, the region in the  $\MA$--$\tb$-plane where the coupling 
of the lightest Higgs boson to down-type fermions is suppressed 
is modified. The presently not calculated higher-order 
corrections to the Higgs-boson mass matrix are estimated to shift the
mass of the lightest Higgs boson by up to $3 \gev$.
\end{abstract}
%\pacs{}

\def\thefootnote{\arabic{footnote}}
\setcounter{page}{0}
\setcounter{footnote}{0}

\newpage

%%%%%%%%%%%%%%%%%%%%%%%%%%%%%%%%%%%%%%%%%%%%%%%%%%%%%%%%%%%%%%
%%%%%%%%%%%%%%%%%%%%%%%%%%%%%%%%%%%%%%%%%%%%%%%%%%%%%%%%%%%%%%
\section{Introduction}

A crucial prediction of the Minimal Supersymmetric Standard Model
(MSSM)~\cite{susy} is the existence of at least one light Higgs
boson. The search for this particle 
is one of the main goals at the present and the next generation
of colliders. Direct searches at LEP have already ruled out a
considerable fraction of the MSSM parameter space, and the forthcoming
high-energy experiments at the Tevatron, the LHC and a future Linear
Collider (LC) will either
discover a light Higgs boson or rule out the MSSM as a viable theory
for physics at the weak scale. Furthermore, if one or more Higgs
bosons are discovered, their masses and couplings will be determined
with high accuracy at a future LC. Thus, a
precise knowledge of the dependence of the masses and mixing angles of
the MSSM Higgs sector on the relevant supersymmetric parameters is of
utmost importance to reliably compare the predictions of the MSSM with
the (present and future) experimental results.

We recall that the Higgs sector of the MSSM~\cite{hhg} consists of two
neutral $\cp$-even Higgs bosons, $h$ and $H$ ($\mh < \mH$), the
$\cp$-odd $A$~boson, and two charged Higgs bosons, $H^\pm$.  At the
tree-level, $\mhtree$ and $\mHtree$ can be calculated in terms of the Standard
Model (SM) gauge couplings and two additional MSSM parameters,
conventionally chosen as $\MA$ and $\tb$, the ratio of the two vacuum
expectation values ($\tb = v_2/v_1$). The two masses are obtained by
rotating the neutral $\cp$-even Higgs boson
mass matrix with an angle $\alpha$, 
\BEA M_{\rm Higgs}^{2, {\rm tree}} &=& \ML \MA^2 \SQb +
\MZ^2 \CQb & -(\MA^2 + \MZ^2) \Sb \Cb \\ -(\MA^2 + \MZ^2) \Sb \Cb &
\MA^2 \CQb + \MZ^2 \SQb \MR ,
\label{higgsmassmatrixtree}
\EEA
with  $\alpha$ satisfying
\BE
\tan 2\alpha = \tan 2\beta \frac{\MA^2 + \MZ^2}{\MA^2 - \MZ^2},
\quad - \frac{\pi}{2} < \alpha < 0 . 
\label{alpha}
\EE

\bigskip
In the Feynman diagrammatic (FD) approach the higher-order corrected 
Higgs boson masses are derived by finding the
poles of the $h,H$-propagator 
matrix whose inverse is given by
\BE
\left(\Delta_{\rm Higgs}\right)^{-1}
= - i \ML p^2 -  \mHtree^2 + \hSi_{HH}(p^2) &  \hSi_{hH}(p^2) \\
     \hSi_{hH}(p^2) & p^2 -  \mhtree^2 + \hSi_{hh}(p^2) \MR,
\label{higgsmassmatrixnondiag}
\EE
where the $\hSi(p^2)$ denote the renormalized Higgs-boson self-energies,
$p$ being the momentum flowing on the external legs.
Determining the poles of the matrix $\Delta_{\rm Higgs}$ in
\refeq{higgsmassmatrixnondiag} is equivalent to solving
the equation
\begin{equation}
\left[p^2 - \mhtree^2 + \hSi_{hh}(p^2) \right]
\left[p^2 - \mHtree^2 + \hSi_{HH}(p^2) \right] -
\left[\hSi_{hH}(p^2)\right]^2 = 0\,.
\label{eq:proppole}
\end{equation}

The status of the available results for the self-energy contributions to
\refeq{higgsmassmatrixnondiag} can be summarized as follows. For the
one-loop part, the complete result within the MSSM is 
known~\cite{ERZ,mhiggsf1lA,mhiggsf1lB,mhiggsf1lC}. The by far dominant
one-loop contribution is the \order{\alt} term due to top and stop 
loops ($\alt \equiv h_t^2 / (4 \pi)$, $h_t$ being the 
superpotential top coupling).
Concerning the two-loop
effects, their computation is quite advanced and it has now reached a
stage such that all the presumably dominant
contributions are known. They include the strong corrections, usually
indicated as \order{\alt\als}, and Yukawa corrections, \order{\alt^2},
to the dominant one-loop \order{\alt} term, as well as the strong
corrections to the bottom/sbottom one-loop \order{\alb} term ($\alb
\equiv h_b^2 / (4\pi)$), i.e.\ the \order{\alb\als} contribution. The
latter can be relevant for large values of $\Tb$. Presently, the
\order{\alt\als}~\cite{mhiggsEP1b,mhiggsletter,mhiggslong,mhiggsEP0,mhiggsEP1},
\order{\alt^2}~\cite{mhiggsEP1b,mhiggsEP3,mhiggsEP2} and the
\order{\alb\als}~\cite{mhiggsEP4} contributions to the self-energies
are known for vanishing external momenta.  In the (s)bottom
corrections the all-order resummation of the $\Tb$-enhanced terms,
\order{\alb(\als\tb)^n}, is also performed \cite{deltamb1,deltamb}.

In this paper we are going to present, in view of the recent
achievements obtained in the knowledge of the two-loop corrections,
updated results for various quantities of physical interest. In our
analysis we employ the latest version of the Fortran code
\fh~\cite{feynhiggs,feynhiggs1.2}, namely \fh1.3, that evaluates the
MSSM neutral $\cp$-even Higgs sector masses and the mixing angle.

The paper is organized as follows: in Section \ref{sec:fhstatus} we 
give a brief summary of recent theoretical improvements in the MSSM
Higgs sector and describe the corresponding modifications in the code
\fh\ in order to comprise all existing higher-order results.
We compare the results for $\mh$ obtained employing different approximations
in the treatment of the two-loop corrections. In Section 
\ref{sec:phenoimp} we present the results for the 
phenomenological implications of our improved knowledge of the two-loop
contributions to the Higgs boson self-energies. Section \ref{sec:hocorr} 
contains a discussion of the presently unknown contributions 
in the prediction for the $\cp$-even Higgs-boson masses of the MSSM,
and an estimate of their possible numerical importance is given.
Finally, in Section \ref{sec:concl} we draw our conclusions.

%%%%%%%%%%%%%%%%%%%%%%%%%%%%%%%%%%%%%%%%%%%%%%%%%%%%%%%%%%%%%%
%%%%%%%%%%%%%%%%%%%%%%%%%%%%%%%%%%%%%%%%%%%%%%%%%%%%%%%%%%%%%%

\section{Recent improvements in the MSSM Higgs sector and implementation
in \fh}
\label{sec:fhstatus}

In order to discuss the impact of recent improvements in the MSSM Higgs
sector we will make use of the program
\fh~\cite{feynhiggs,feynhiggs1.2}, which 
is a Fortran code for the evaluation
of the neutral $\cp$-even Higgs sector of the MSSM including
higher-order corrections to the renormalized Higgs boson
self-energies. The original release included the well known full
\onel\ corrections~\cite{ERZ,mhiggsf1lA,mhiggsf1lB,mhiggsf1lC}, the
two-loop leading, momentum-independent, \order{\alt\als} correction
in the top/stop sector~\cite{mhiggsletter,mhiggslong,mhiggsEP1} as
well as the two-loop leading logarithmic corrections at
\order{\alt^2}~\cite{mhiggsRG1,mhiggsRG2}.  By two-loop
momentum-independent corrections here and hereafter we mean the
two-loop contributions to Higgs boson self-energies evaluated at zero
external momenta. At the one-loop level, the momentum-independent
contributions are the dominant part of the self-energy corrections, that, 
in principle, should be 
evaluated at external momenta equal to the poles of the
$h,H$-propagator matrix, \refeq{higgsmassmatrixnondiag}.

The new version of \fh\  contains a modification in the one-loop part
due to a different renormalization prescription employed 
and includes also several new corrections to the Higgs boson self-energies
that have recently been calculated. These changes are described in the 
following subsections.  

With the implementation of the latest results obtained in the MSSM Higgs
sector, \fh\ comprises all available higher-order corrections and thus
allows the presently most precise prediction of the masses of the
$\cp$-even Higgs bosons and the corresponding mixing angle.
The latest version of \fh, \fh1.3, can be obtained from 
{\tt www.feynhiggs.de}.

%%%%%%%%%%%%%%%%%%%%%%%%%%%%%%%%%%%%%%%%%%%%%%%%%%%%%%%%%%%%%%
%%%%%%%%%%%%%%%%%%%%%%%%%%%%%%%%%%%%%%%%%%%%%%%%%%%%%%%%%%%%%%

\subsection{Hybrid renormalization scheme at the \onel\ level}
\label{subsec:newren}

\fh\ is based on the FD approach with on-shell renormalization 
conditions~\cite{mhiggslong}. This means in particular that all the 
masses in the FD result are the physical ones, i.e.\ they correspond to
physical observables. Since \refeq{eq:proppole} is solved
iteratively, the result for $m_h$ and $m_H$ contains a dependence on
the field-renormalization constants of $h$ and $H$, which is 
formally of higher order. Accordingly, there is some freedom in choosing 
appropriate renormalization conditions for fixing the
field-renormalization constants (this can also be interpreted as
affecting the 
renormalization of $\tb$). Different renormalization conditions have
been considered in the literature, e.g.\ ($\hSip$ denotes the derivative 
with respect to the external momentum squared):
\begin{enumerate}
\item
on-shell renormalization for $\hSi_Z, \hSi_A,
\hSip_A, \hSi_{AZ}$, and  
$\de v_1/v_1 = \de v_2/v_2$~\cite{mhiggsf1lC}
\item
on-shell renormalization for 
$\hSi_Z, \hSi_A, \hSi_{AZ}$, and 
$\de v_i = \de v_{i, {\rm div}}, i = 1,2$~\cite{mhiggsf1lB,eehZhA}
\item
on-shell renormalization for $\hSi_Z, \hSi_A$~,
\drbar\ renormalization (employing dimensional reduction~\cite{dred})
for $\de Z_h, \de Z_H$ and $\tb$~\cite{feynhiggs1.2}.
\end{enumerate}
Versions of \fh\ previous than the 1.2 release were based on type-1 
renormalization conditions, thus requiring the derivative of the $A$~boson
self-energy.  The latest versions  employ instead the hybrid
\drbar/on-shell type 3 conditions\cite{feynhiggs1.2}. The choice of a 
\drbar\ definition%
\footnote{
A minimal definition for $\Tb$ and the
field renormalization constants was already employed in
\citere{mhiggsf1lA}, in the analysis of the one-loop (s)top--(s)bottom
contribution to the Higgs boson self-energies.
}%
~for $\de Z_h, \de Z_H$ and $\tb$ requires to specify a renormalization
scale $Q^2$ at which these parameters are defined. This scale is set
to $\mt^2$ in \fh, but can be changed by the user. These new
renormalization conditions lead to a more stable behavior around
thresholds, e.g.\ $\MA= 2\, \mt$, and avoid unphysically large
contributions in certain regions of the MSSM parameter space (a
detailed discussion can be found in \citere{mhiggsrenorm}; see also
\citere{ayresdoink}).

%%%%%%%%%%%%%%%%%%%%%%%%%%%%%%%%%%%%%%%%%%%%%%%%%%%%%%%%%%%%%%
%%%%%%%%%%%%%%%%%%%%%%%%%%%%%%%%%%%%%%%%%%%%%%%%%%%%%%%%%%%%%%
\subsection{Two-loop \order{\alt^2} corrections}
\label{subsec:nonlogaltsq}

In the previous version of \fh\ the two-loop \order{\alt^2}
corrections were implemented in a simplified form, taken over from the
result of \cite{mhiggsRG1,mhiggsRG2}, obtained with the
renormalization group (RG) method.  The latter result included the
two-loop leading logarithmic corrections, proportional to the square
of $t \equiv \log\,(\msusy^2/m_t^2)$ (where $\msusy$ is a common scale
for the soft SUSY-breaking parameters), while the next-to-leading
logarithmic terms, linear in $t$, were not totally accounted for.

Recently, the two-loop \order{\alt^2} corrections in the limit of
zero external momentum became available, first only for the lightest
eigenvalue, $\mh$, and in the limit $\MA \gg \MZ$~\cite{mhiggsEP3},
then for all the entries of the Higgs propagator matrix for arbitrary
values of $\MA$~\cite{mhiggsEP2}. They were obtained in the
effective-potential approach, that allows to construct the Higgs
boson self-energies, at zero external momenta, by taking the relevant
derivatives of the field-dependent potential.  In this procedure it
is important, in order to make contact with the physical $\MA$, to
compute the effective potential as a function of both $\cp$-even and
$\cp$-odd fields, as emphasized in \citere{mhiggsEP1}. In the
evaluation of the \order{\alt^2} corrections, the specification of a
renormalization prescription for the Higgs mixing parameter $\mu$ is also
required and it has been chosen as \drbar.
In \fh1.3, which includes the complete two-loop \order{\alt^2} corrections, 
the corresponding renormalization scale is fixed to be the same as for 
$\de Z_h, \de Z_H$ and $\tb$. 

%%%%%%%%%%%%%%%%%%%%% FIGURE %%%%%%%%%%%%%%%%%%%%%%%%%%%%%%%%%%%%%%%%%%
\begin{figure}[t]
\begin{center}
\epsfig{figure=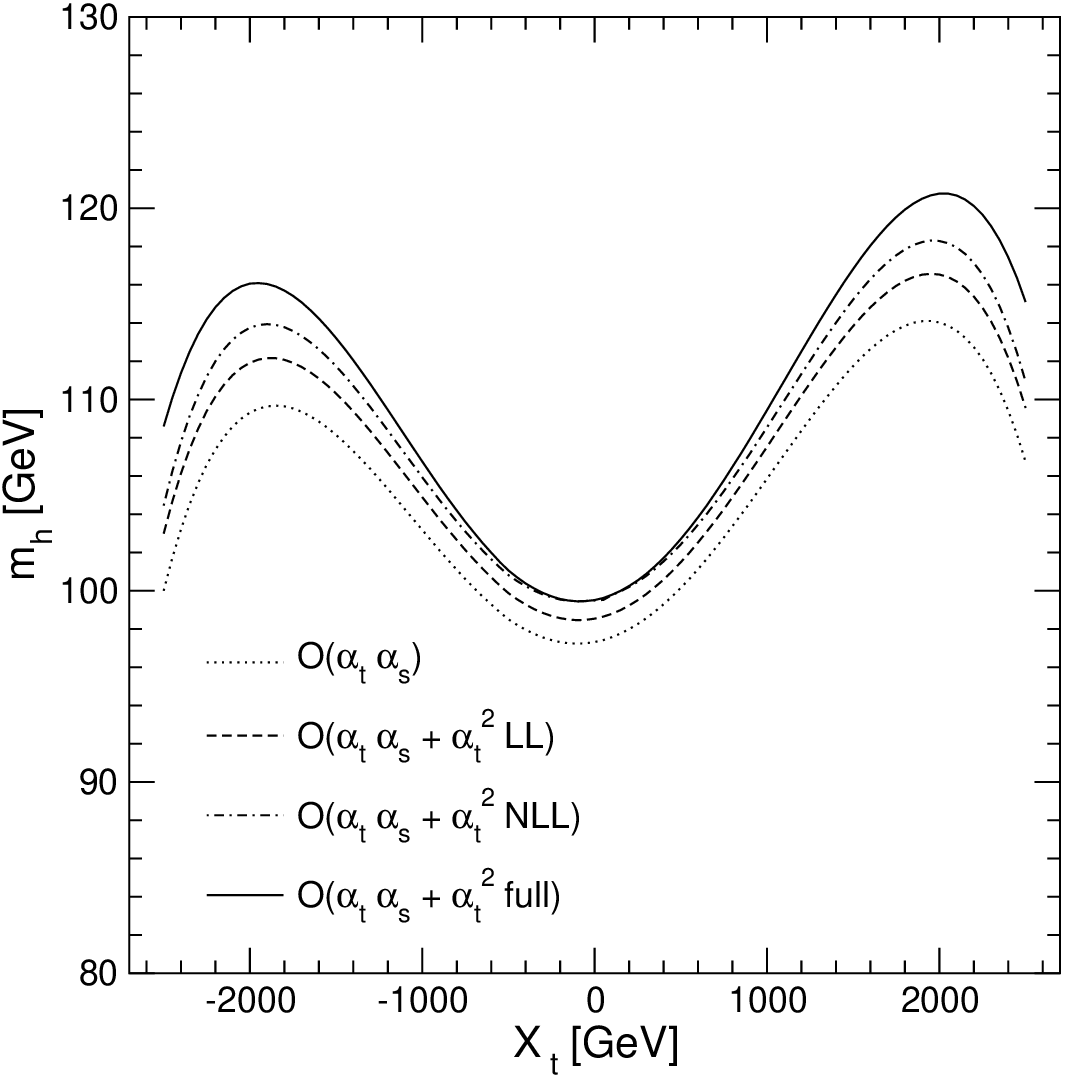,width=12cm}
\end{center}
\caption[]{ Two-loop corrected
$\mh$ as a function of $\Xt$ in various steps of approximation.
The relevant MSSM parameters are chosen as $\tb=3\,,\, 
\mt^{\rm pole} = 174.3 \gev$,
$\MstL = \MstR = M_A = \mu = 1 \tev$ and $\mgl = 800 \gev$.
The meaning of the different curves is explained in the text.
}
\label{fig:RGmeth}
\end{figure}
%%%%%%%%%%%%%%%%%%%%% FIGURE %%%%%%%%%%%%%%%%%%%%%%%%%%%%%%%%%%%%%%%%%%
 
The availability of the complete result for the momentum-independent
part of the \order{\alt^2} corrections allows us to judge the quality
of results that incorporate only the logarithmic contributions. 
In \reffi{fig:RGmeth} we plot the two-loop corrected $\mh$ as a
function of the stop mixing parameter, $X_t \equiv \At -\mu/\tb$,
where $\At$ denotes the trilinear soft SUSY breaking Higgs--stop
coupling. We use the following convention for the stop mass matrix,
\BE
\label{stopmassenmatrix}
{\cal M}^2_{\Stop} =
  \ML \MstL^2 + \mt^2 + \CZb (\edz - \frac{2}{3} \sw^2) \MZ^2 &
      \mt \Xt \\
      \mt \Xt &
      \MstR^2 + \mt^2 + \frac{2}{3} \CZb \sw^2 \MZ^2
  \MR~.
\EE
For simplicity, the soft SUSY breaking parameters in the
diagonal entries of the stop mass matrix, $\MstL, \,\MstR$, are chosen
to be equal, $\MstL = \MstR = \msusy$. For the numerical analysis
  $\msusy$ as well as
$\MA$ and $\mu$, are chosen to be all equal to $1 \tev$, while the
gluino mass is $\mgl = 800 \gev$, and $\tb=3$. If not otherwise stated, in
all plots below we choose the trilinear couplings in the stop and
sbottom sectors equal to each other, $\Ab = \At$, and set $M_2 =
\msusy$, where $M_2$ is the SU(2) gaugino mass parameter (the U(1)
gaugino mass parameter is obtained via the GUT relation, 
$M_1 = 5/3\, \sw^2/\cw^2\, M_2$).

The solid and dotted lines in \reffi{fig:RGmeth} 
are computed with and without the inclusion of the full
\order{\alt^2} corrections, while the dashed and dot--dashed ones are
obtained including only the logarithmic contributions. In particular,
the dashed curve corresponds to the result of \cite{mhiggsRG1},
obtained with the one-loop
renormalization group method. The dot--dashed curve,
instead, corresponds to the leading and next-to-leading logarithmic
terms that, within this simplifying choice of the MSSM parameters, can
be easily singled out from the complete \order{\alt^2} result. As
mentioned above, the two approximate results agree with each other in
the terms proportional to $t^2$, whereas they disagree in the terms
linear in $t$, giving rise to the difference between the dashed and
dot--dashed curves.  In fact, as long as one is only concerned about
the leading logarithmic terms, it is correct to use in the
renormalization group equations (RGE) for the Higgs quartic couplings
the one-loop $\be$-function as is done in
\citere{mhiggsRG1}. Moreover, the exact definition of the various
parameters in the one-loop part is not important. Instead, when the
next-to-leading logarithmic terms are examined the \twol\
$\be$-function has to be employed, giving rise to additional single
logarithmic terms at the two-loop order. Furthermore, the results
from \cite{mhiggsRG1} are valid under the assumption that the
one-loop part of the corrections is written in terms of running SM
parameters in the \msbar\ renormalization scheme, whereas the
one-loop computation in \fh\ employs on-shell parameters. As
discussed in \cite{mhiggsRG2,mhiggsEP0,reconcA,reconciling}, this amounts to a
shift in the two-loop corrections that also affects the
next-to-leading logarithmic part of the \order{\alt^2} results.

From \reffi{fig:RGmeth} it can also be seen that, for small $X_t$, the
full \order{\alt^2} result is very well reproduced by the logarithmic
approximation, once the next-to-leading terms are correctly taken
into account. On the other hand, when $X_t$ is large there are
significant differences, amounting to several GeV, between the
logarithmic approximation and the full result. Such differences are
due to non-logarithmic terms that scale like powers of $\Xt/\msusy$. 
It should be noted that for more general choices of the
MSSM parameters the renormalization group method becomes rather
involved (see e.g.\ \cite{espnav}, where the case of a large splitting
between $\MstL$ and $\MstR$ is discussed), and a suitable
next-to-leading logarithmic
approximation to the full result is much more difficult to devise.

%%%%%%%%%%%%%%%%%%%%%%%%%%%%%%%%%%%%%%%%%%%%%%%%%%%%%%%%%%%%%%
%%%%%%%%%%%%%%%%%%%%%%%%%%%%%%%%%%%%%%%%%%%%%%%%%%%%%%%%%%%%%%

\subsection{Two-loop sbottom corrections}
\label{subsec:abass}
Due to the smallness of the bottom mass, the \order{\alb} one-loop
corrections to the Higgs boson self-energies can be numerically
non-negligible only for $\tb \gg 1$ and sizable values of the $\mu$
parameter.  In fact, at the classical level $h_b / h_t = (m_b / m_t)
\tb$, thus $\tb \gg 1$ is needed in order to have $\alb \sim \alt$
in spite of $m_b \ll m_t$.  In contrast to the \order{\alt}
corrections where both top and stop loops give sizable contributions,
in the case of
the \order{\alb} corrections the numerically dominant contributions
come from sbottom loops: those coming from bottom loops are
always suppressed by the small value of the bottom mass. A sizable
value of $\mu$ is then required to have sizable sbottom--Higgs scalar
interactions in the large-$\tb$ limit.

The relation between the bottom-quark mass and the Yukawa coupling
$h_b$, which controls also the interaction between the Higgs fields and
the sbottom squarks, reads at lowest order $m_b =h_b v_1 /
\sqrt{2}$. This relation is affected at one-loop order by large radiative
corrections \cite{deltamb1}, proportional to $h_b v_2$, 
giving rise in general to $\tb$-enhanced contributions.
These terms proportional to $v_2$, often indicated as threshold
corrections to the bottom mass, are generated either by
gluino--sbottom one-loop diagrams, resulting in \order{\alb\al_s}
corrections to the Higgs masses, or by chargino--stop ones, giving
\order{\alb\alt} corrections.  Because the $\tb$-enhanced
contributions can be numerically relevant, an accurate determination
of $h_b$ from the experimental value of the bottom mass requires a
resummation of such effects to all orders in the perturbative
expansion, as described in \citere{deltamb}.

Concerning the sbottom corrections to the renormalized Higgs boson
self-energies, the version 1.2.2 of \fh\ included the full one-loop
contribution, improved by the resummation of the $\tb$-enhanced
effects in the relation between $h_b$ and $m_b$ according to the
effective Lagrangian formalism developed in \citere{deltamb}. This
takes into account the \order{\alb(\als\tb)^n} corrections to all
orders in $n$. Numerically this is by far the dominant part of the
contributions from the sbottom sector.

%%%%%%%%%%%%%%%%%%%%% FIGURE %%%%%%%%%%%%%%%%%%%%%%%%%%%%%%%%%%%%%%%%%%
\begin{figure}[t]
\begin{center}
\epsfig{figure=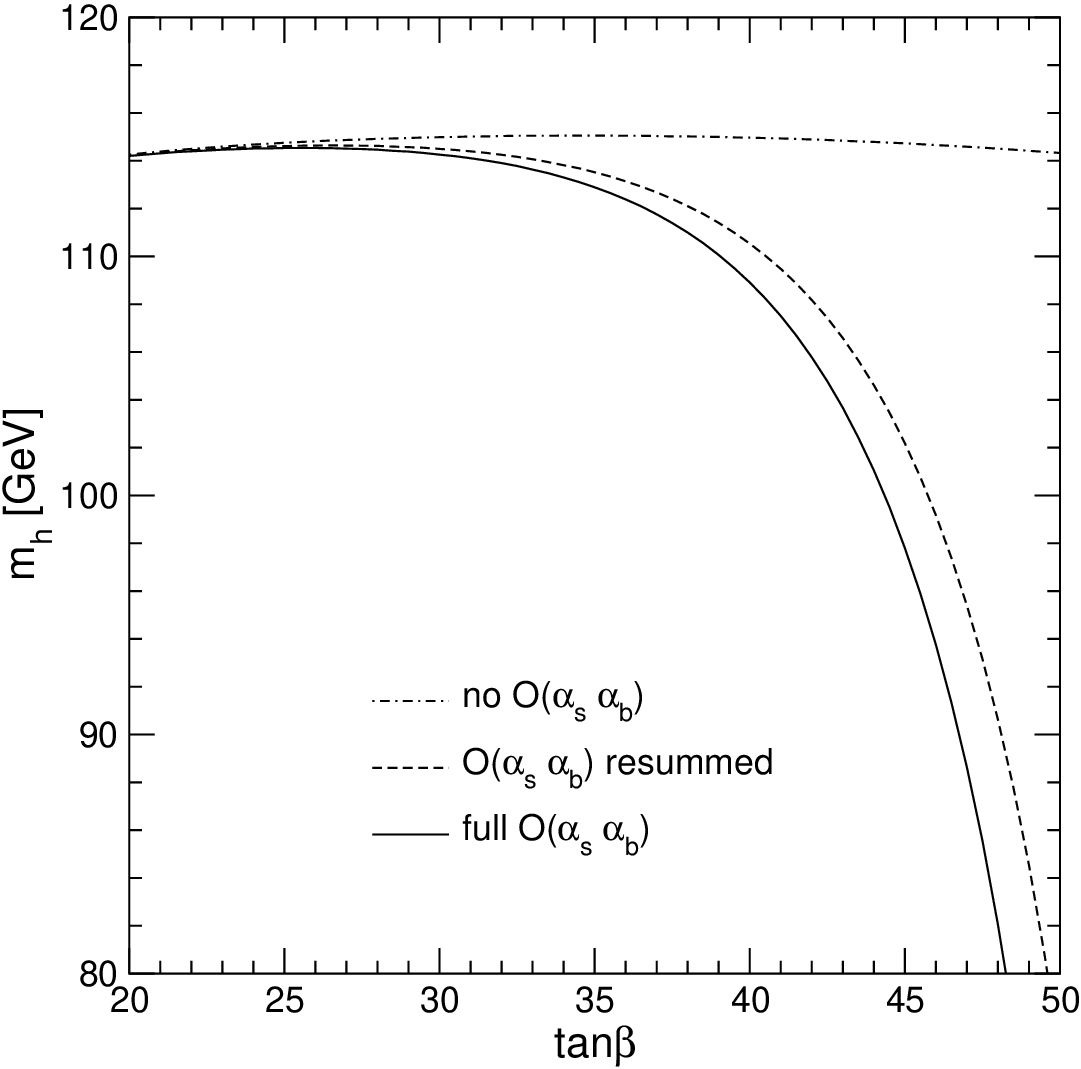,width=12cm}
\end{center}
\caption[]{ The result for the lightest $\cp$-even Higgs-boson mass in 
the MSSM, $\mh$, as obtained with the program \fh\ is shown 
as a function of $\tb$ for $\MA = 120$ GeV, $\mu = -1$~TeV, 
$\MstL = \MstR = \MsbR = \mgl = 1$ TeV, $A_t = A_b = 2$~TeV. 
The meaning of the different curves is explained in the text.}
\label{fig:sbottom}
\end{figure}
%%%%%%%%%%%%%%%%%%%%% FIGURE %%%%%%%%%%%%%%%%%%%%%%%%%%%%%%%%%%%%%%%%%%

Very recently, the complete two-loop, momentum-independent,
\order{\alb\al_s} corrections (which are not included in the
\order{\alb(\als\tb)^n} resummation) have been
computed~\cite{mhiggsEP4}.  The result obtained makes use of an
appropriate choice of renormalization conditions on the relevant
parameters that allows to disentangle the genuine two-loop effects
from the large threshold corrections to the bottom mass, and also ensures
the decoupling of heavy gluinos. The complete momentum-independent
\order{\alb\al_s}
corrections have now also been implemented in \fh1.3.

To appreciate the importance of the various sbottom contributions, we
plot in Fig.~\ref{fig:sbottom} the light Higgs mass $\mh$ as a
function of $\tb$. The SM running bottom mass computed at the top mass
scale, $m_b(m_t) = 2.74$ GeV, is used in order to account for the
universal large QCD corrections.  The relevant MSSM parameters are
chosen as $\MA = 120$~GeV, $\mu = -1$~TeV, $\MstL = \MstR = \MsbR =
\mgl = 1$~TeV, $A_t = A_b = 2$~TeV (where $\MsbR$ is a soft
SUSY breaking parameter in the sbottom sector). The dot--dashed curve
in Fig.~\ref{fig:sbottom} includes the full one-loop contribution as
well as the two-loop \order{\alt\als + \alt^2} corrections (the
latter being approximately $\tb$-independent when $\tb$ is
large). The dashed curve includes also the resummation of the
$\tb$-enhanced threshold effects in the relation between $h_b$ and
$m_b$. Finally, the solid curve includes in addition the complete
\order{\alb\al_s} two-loop corrections of \cite{mhiggsEP4}. In the
last two curves, the steep dependence of $\mh$ on $\tb$ when the
latter is large is driven by the sbottom contributions. We see that,
although the $\tb$-enhanced threshold effects account for the bulk of
the sbottom contributions beyond one-loop, the genuine
\order{\alb\al_s} two-loop corrections can still shift $\mh$ by
several GeV for large values of $\tb$ and $\mu$.%
\footnote{
In the region of large $\tb$ the bottom Yukawa coupling receives large
corrections, see \refeq{effcoupling} below. For the parameter range
shown in Fig.~\ref{fig:sbottom} the quantity $\De_b$, defined in
\refeq{effcoupling}, is negative but does not get close to $-1$,
$|\De_b| \lsim 0.6$. Thus, the perturbative treatment in
Fig.~\ref{fig:sbottom} seems to be justified.
}%
%

%%%%%%%%%%%%%%%%%%%%%%%%%%%%%%%%%%%%%%%%%%%%%%%%%%%%%%%%%%%%%%
%%%%%%%%%%%%%%%%%%%%%%%%%%%%%%%%%%%%%%%%%%%%%%%%%%%%%%%%%%%%%%

\section{Phenomenological implications}
\label{sec:phenoimp}

The improved knowledge of the two-loop contributions to the Higgs-boson
self-energies results in a very precise prediction for the Higgs-boson 
masses and mixing angle with interesting implications for MSSM parameter
space analyses. In this section we are going to discuss 
possible implications on the upper limit
on $\mh$ within the MSSM and on the corresponding limit on $\tb$ arising
from confronting the upper bound on $\mh$ with the lower limit from
Higgs searches. We furthermore investigate
the modifications that are
induced in the couplings of the lightest Higgs boson to the down-type
fermions.

%%%%%%%%%%%%%%%%%%%%%%%%%%%%%%%%%%%%%%%%%%%%%%%%%%%%%%%%%%%%%%
%%%%%%%%%%%%%%%%%%%%%%%%%%%%%%%%%%%%%%%%%%%%%%%%%%%%%%%%%%%%%%

\subsection{Limits on $\mh$ and $\tb$}
\label{subsec:mhtblimits}
The theoretical upper bound on the lightest Higgs-boson mass as a
function of $\tb$ can be combined with the results from direct
searches at LEP to constrain $\tb$. The diagonalization of the
tree-level mass matrix, \refeq{higgsmassmatrixtree}, yields a value for
$\mhtree$ that is maximal when $\MA \gg \MZ$, in which case 
$\mhtree^2 \simeq \MZ^2\, \CQZb$, 
which vanishes for $\tb=1$. 
Radiative corrections significantly increase the light
Higgs-boson mass compared to its tree-level value, but still $\mh$ is
minimized for values of $\tb$ around one. Thus, in principle, the region
of low $\tb$ can be probed experimentally via the search for the
lightest MSSM Higgs boson~\cite{LEPHiggs}. If the remaining MSSM
parameters are tuned in such a way to obtain the maximal value of $\mh$
as a function of $\tb$ (for reasonable values of $\msusy$ and taking
into account the experimental uncertainties of $\mt$ and the
other SM input parameters as well as the theoretical uncertainties from
unknown higher-order corrections), the experimental lower bound on
$\mh$ can be used to obtain exclusion limits for $\tb$.
While in general a detailed investigation of a variety of different
possible production and decay modes is necessary in order to determine
whether a particular point of the MSSM parameter space can be excluded
via the Higgs searches or not, the situation simplifies considerably in
the region of small $\tb$ values. In this parameter region the lightest
$\cp$-even Higgs boson of the MSSM couples to the $Z$~boson with
SM-like strength, and its decay into a $b\bar b$ pair is not significantly
suppressed. Thus, within good approximation, constraints on $\tb$ can 
be obtained in this parameter region by confronting the exclusion bound
on the SM Higgs boson with the upper limit on $\mh$ within the MSSM.
We use this approach below in order to discuss the implications of the
new $\mh$ evaluation on $\tb$ exclusion bounds.

Concerning the upper bound on $\mh$ within the MSSM,
the one-loop corrections contribute positively to $\mh^2$. The
two-loop effects of \order{\alt\als} and \order{\alt^2}, on the other
hand, enter with
competing signs, the former reducing $\mh^2$ while the latter giving a
(smaller) positive contribution.  The actual bound that can be derived depends
sensitively on the precise value of the top-quark mass, because the
dominant one-loop contribution to $\mh^2$, as well as the
two-loop \order{\alt\als} term, scale as $\mt^4$.  Furthermore, a
large top mass amplifies the relative importance of the two-loop
\order{\alt^2} correction, because of the additional $\mt^2$ factor.

In order to discuss restrictions on the MSSM parameter space it has
become customary in the recent years to refer to so-called benchmark
scenarios of MSSM parameters~\cite{benchmark,sps}. The $\mhmax$ benchmark
scenario~\cite{benchmark} has been designed such that for fixed values
of $\mt$ and $\msusy$ the predicted value of the lightest $\cp$-even
Higgs-boson mass is maximized for each value of $\MA$ and $\tb$. The
value of the 
top-quark mass is fixed to its experimental central value, $\mt = 174.3
\gev$, while the SUSY parameters are taken as (referring to their
on-shell values according to the FD result as implemented in \fh):
\BEA 
&& \msusy = 1\tev, \quad \mu = -200 \gev, \quad 
                                M_2 = 200 \gev, \non \\ 
&& \Xt^{\rm FD}  = 2\, \msusy, \quad \Ab = \At, \quad 
   \mgl = 0.8\,\msusy~,
\label{mhmax} 
\EEA 
and we will investigate below the case $\MA = 1$~TeV.

%%%%%%%%%%%%%%%%%%%%% FIGURE %%%%%%%%%%%%%%%%%%%%%%%%%%%%%%%%%%%%%%%%%%
\begin{figure}[t]
\begin{center}
\epsfig{figure=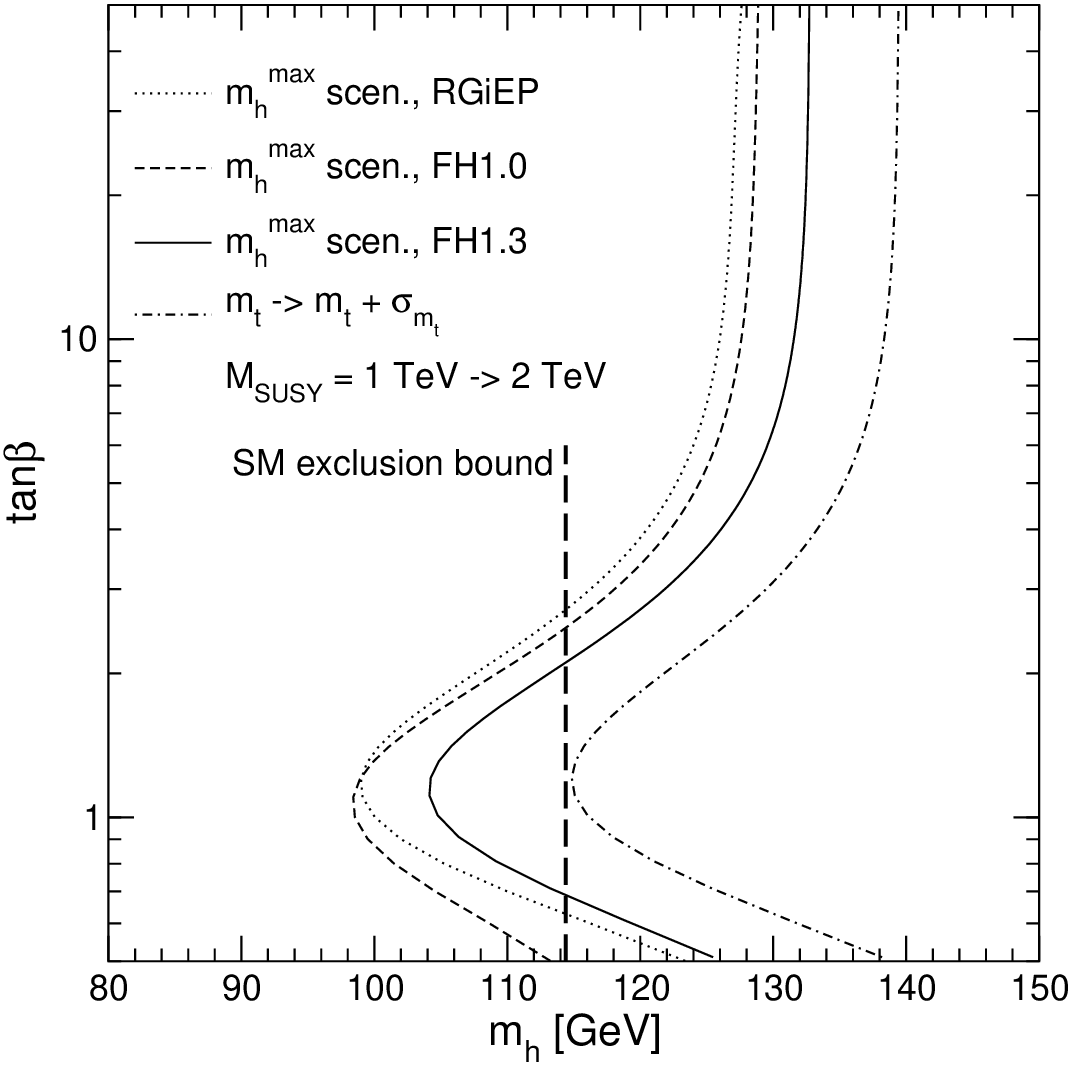,width=12cm}
\end{center}
\vspace{-1em}
\caption[]{The result for the lightest $\cp$-even Higgs-boson mass, $\mh$,
as a function of $\tb$ in the $\mhmax$~scenario. The
dotted curve has been obtained with a renormalization-group improved
effective potential
method, using the code {\em subhpoledm}.  The dashed curve
corresponds to the result obtained with \fh1.0, while the full curve shows the 
result of \fh1.3, where the improvements described in
\refse{sec:fhstatus} are included.
The dot-dashed curve, also 
obtained with \fh1.3, employs $\mt = 179.4 \gev$ and $\msusy = 2
\tev$. The vertical long-dashed line corresponds to the LEP exclusion
bound for the SM Higgs boson of $114.4 \gev$.  }
\label{fig:mhtb}
\end{figure}
%%%%%%%%%%%%%%%%%%%%% FIGURE %%%%%%%%%%%%%%%%%%%%%%%%%%%%%%%%%%%%%%%%%%

In \reffi{fig:mhtb} we plot $\mh$ as a function of $\tb$ in the
$\mhmax$~scenario. The dashed and solid curves correspond to the
result obtained with the previous (used for the LEP evaluations so
far~\cite{LEPHiggs}) and the latest (which will be used for the final
LEP evaluations~\cite{arnulf}) versions of \fh, respectively. The two
versions, \fh1.0 and \fh1.3, differ by the recent improvements obtained
in the MSSM Higgs sector which are described in \refse{sec:fhstatus}.
For comparison, also the result obtained with a renormalization-group
improved effective potential method is indicated. The dotted curve in
Fig.~\ref{fig:mhtb} corresponds to the code 
{\em subhpoledm}~\cite{mhiggsRG1,reconciling,deltamb} in the 
$\mhmax$~scenario, for a $\Stop$~mixing parameter $\Xt^{\msbarm} =
\sqrt{6}\,\msusy$~\cite{mhiggsRG1,reconciling}. It deviates from the
result of \fh1.0 by typically not more than 1~GeV for $\tb \geq 1$.
The LEP exclusion bound for the mass of a SM-like Higgs~\cite{LEPHiggsSM},
$\MHSM \ge 114.4 \gev$, is shown in the figure as a vertical long--dashed 
line. As can be seen from the figure, the improvements on the
theoretical prediction described in \refse{sec:fhstatus}, in particular
the inclusion of the complete momentum-independent \order{\alt^2}
corrections into \fh, gives rise to a significant increase in the upper
bound on $\mh$ as a function of $\tb$. Comparison of this prediction 
with the exclusion bound on a SM-like Higgs shows that the lower limit
on $\tb$ is considerably weakened.

Concerning the interpretation of the results shown in
Fig.~\ref{fig:mhtb}, it should be kept in mind that in the $\mhmax$
benchmark scenario $\mt$ and $\msusy$ are kept fixed, and no theoretical
uncertainties from unknown higher-order corrections are taken into
account. In order to arrive at a more general exclusion bound on $\tb$
that is not restricted to a particular benchmark scenario, the impact of
the parametric and higher-order uncertainties in the prediction for
$\mh$ has to be considered~\cite{tbexcl}. In order to demonstrate in 
particular the dependence of the $\tb$ exclusion bound on the chosen
value of the top pole mass, the dot--dashed curve in Fig.~\ref{fig:mhtb}
shows the result obtained with \fh1.3 where the top-quark mass has been
increased by one standard deviation, $\sigma_{\mt} = 5.1\, \gev$, to
$\mt = 179.4 \gev$, and $\msusy$ has been changed from 1~TeV to 2~TeV.
It can be seen that in this more general scenario no lower limit on
$\tb$ from the LEP Higgs searches can be obtained. 

Constraints from the Higgs searches at LEP do of course play an
important role in regions of the MSSM parameter space where the
parameters are such that $\mh$ does not reach its maximum value. Also in
this case, however, the remaining theoretical uncertainties from unknown
higher-order corrections (see \refse{sec:hocorr} below) have to be
taken into account in order to obtain conservative exclusion limits.

%%%%%%%%%%%%%%%%%%%%%%%%%%%%%%%%%%%%%%%%%%%%%%%%%%%%%%%%%%%%%%
%%%%%%%%%%%%%%%%%%%%%%%%%%%%%%%%%%%%%%%%%%%%%%%%%%%%%%%%%%%%%%

\subsection{Higgs couplings to fermions}
\label{subsec:aeff}

The tree-level couplings of the lightest $\cp$-even Higgs 
boson to the up-type and down-type SM fermions read, respectively:
\BE
\Ga_h^{\,u} = \frac{i \,e \,m_u}{2 \,\sw \,\MW} \, \frac{-\Ca}{\Sb}\,,  
\hspace{2cm}
\Ga_h^{\,d} = \frac{i\, e \,m_d}{2 \,\sw\, \MW}  \, \frac{\Sa}{\Cb}\,,
\label{treecouplings}
\EE
where the factors involving $\al$ and $\be$ reflect the changes in the
MSSM compared to the SM couplings. In the limit of $\MA \gg \MZ$, 
$\al \sim \be - \pi/2$ so that the SM limit is recovered. 
For the lightest $\cp$-even Higgs boson the decay to two bottom quarks
is usually the main decay mode, while the decay to $\tau$~leptons has
usually the second largest branching ratio. However, these channels
are also most significantly affected by loop corrections, which can change 
the situation described above quantitatively and even qualitatively. The
two main sources (besides the SM QCD corrections) are
the Higgs-boson propagator corrections and the corrections modifying
the relation between the bottom quark or $\tau$~lepton mass and the
corresponding Yukawa couplings.

Concerning the former, it has been shown
analytically~\cite{eehZhA,hff} that the momentum-independent
contributions coming from the Higgs-boson propagator corrections can
be incorporated by replacing the tree-level angle $\al$ in
\refeq{treecouplings} with an effective angle $\aeff$, which
diagonalizes the Higgs-boson mass matrix including the self-energy
corrections evaluated at zero external momentum.  Due to the effect of
the higher-order corrections, $\aeff \approx 0$ is possible, i.e.\ the
coupling of the lightest Higgs boson to the down-type SM fermions can
vanish.

The other potentially large corrections to the $\Ga_h^{\,d}$ couplings
come from the $\tb$-enhanced threshold effects in the relation
between the down-type fermion mass and the corresponding Yukawa
coupling~\cite{deltamb1}, already mentioned in
\refse{subsec:abass}. A simple way to take into account these
effects is to employ the effective lagrangian formalism of
\citeres{deltamb,deltamb2}, where the coupling of the lightest Higgs
boson to down-type fermions 
(expressed through the fermion mass) 
is modified according to 
\BE
\left(\Ga_h^{\,d}\right)_{\rm eff} 
= \frac{i\, e \, m_d}{2 \,\sw\, \MW}  \, \frac{\sin \aeff}{\Cb}\,
\left[1 - \frac{\Delta_d}{1+\Delta_d}\,(1+\cot\aeff \,\cot\beta)\right]\, ,
\label{effcoupling}
\EE
where $\Delta_d$ contains the $\tb$-enhanced terms, and other
subleading (i.e.\ non $\tb$-enhanced) corrections have been
omitted. In the case of the coupling to the bottom quarks, the leading
contributions to $\Delta_b$ are of \order{\als} and \order{\alt},
coming from diagrams with sbottom--gluino and stop--chargino loops,
respectively. In the case of the $\tau$ leptons, the leading
contributions are of \order{\alpha_{\tau}}, coming from
sneutrino--chargino loops. The terms containing $\Delta_d$ in
\refeq{effcoupling} may be relevant for large $\tb$ and moderate
values of $\MA$. When $\MA$ is much bigger than $\MZ$, the product
$\cot\aeff \,\cot\beta$ tends to $-1$, and the SM limit is again
recovered.

%%%%%%%%%%%%%%%%%%%%% FIGURE %%%%%%%%%%%%%%%%%%%%%%%%%%%%%%%%%%%%%%%%%%
\begin{figure}[ht!]
\begin{center}
\hspace{-.4cm}
\epsfig{figure=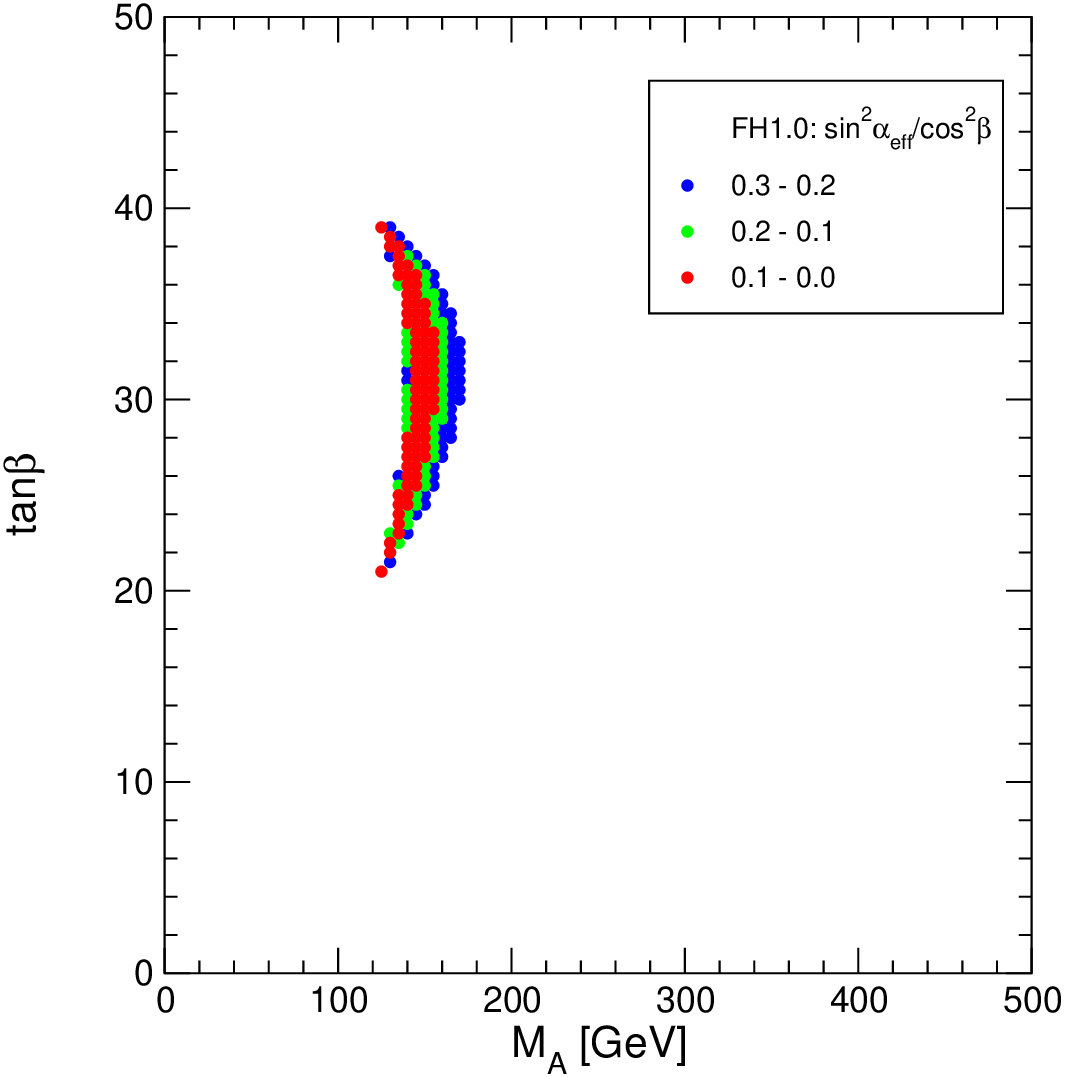,width=8cm,height=8cm}
\hspace{2mm}
\epsfig{figure=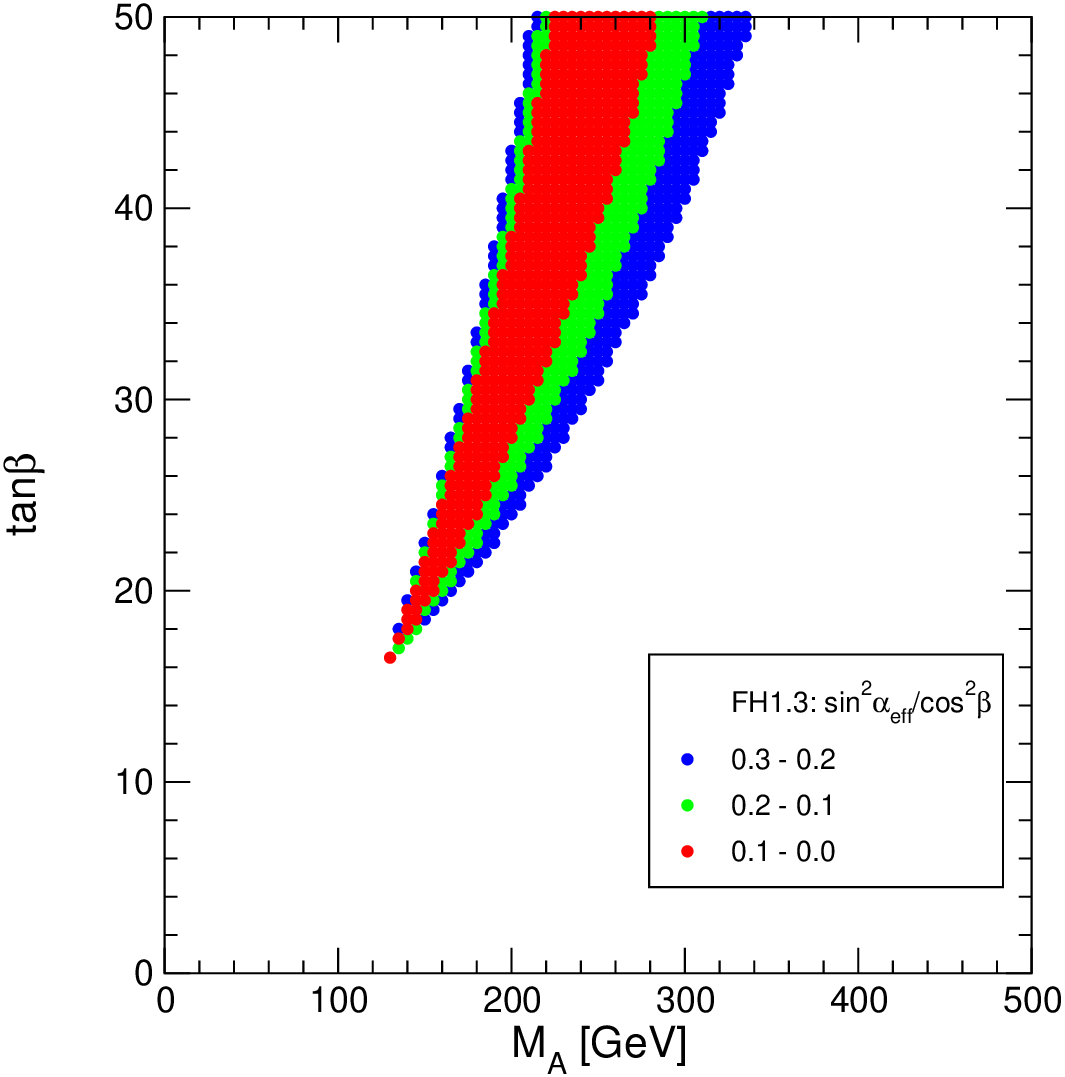,width=8cm,height=8cm}\\[1.0em]
\epsfig{figure=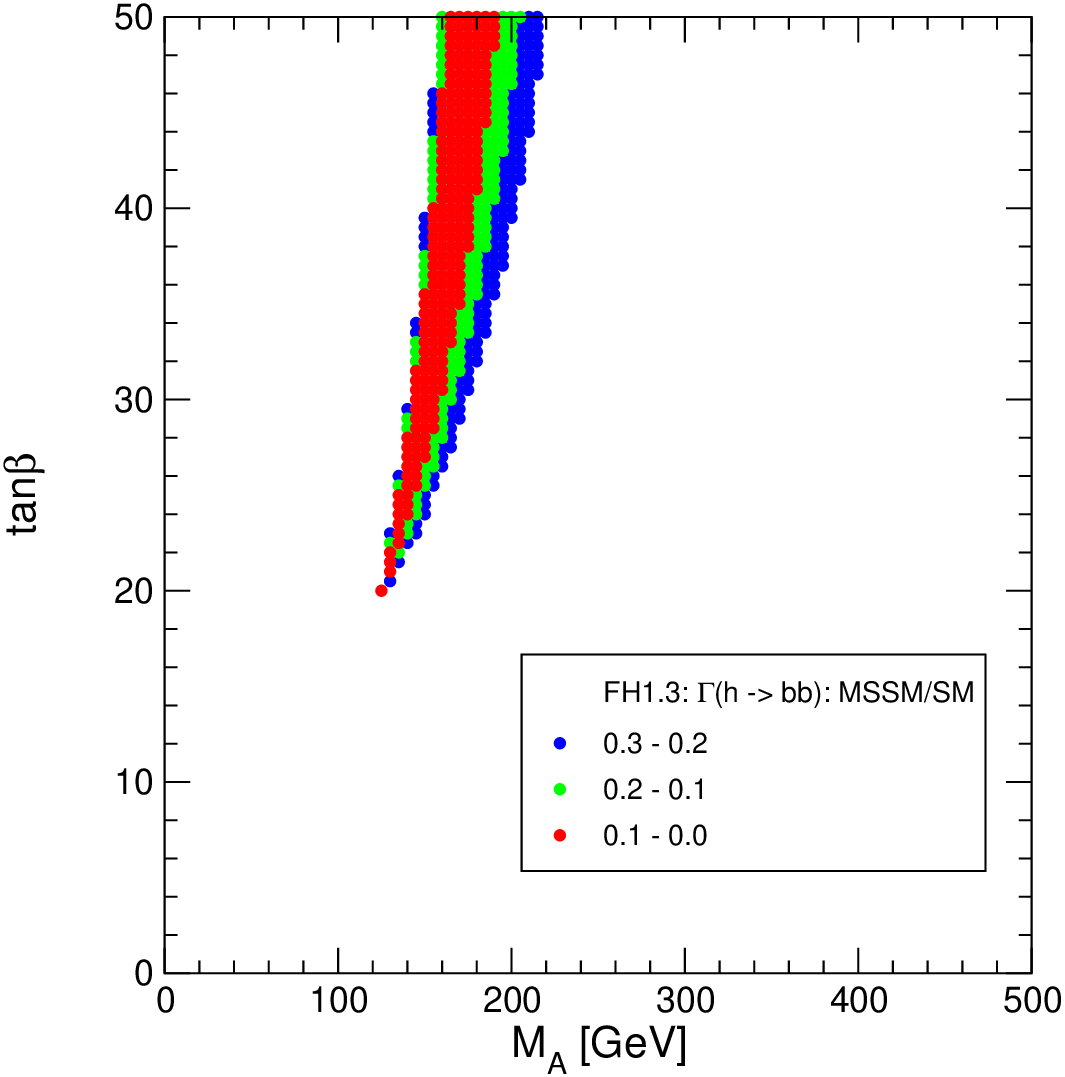,width=8cm,height=8cm}
\end{center}
\vspace{-0.5cm}
\caption{Regions of significant suppression of the coupling of $h$ to 
down-type fermions in the $\MA$--$\tb$-plane within the ``small $\aeff$''
benchmark scenario. The upper left plot shows the ratio 
$\sin^2 \aeff/\CQb$ as evaluated with \fh1.0 (i.e.\ without 
\order{\alb\als} and \order{\alb (\als\tb)^n} corrections), while the
upper right plot shows the same quantity as evaluated with \fh1.3 (i.e.\
including these corrections). The plot in the lower row shows the ratio
$\Gamma(h \rightarrow b\bar{b})_{\rm MSSM}
\,/\,\Gamma(h\rightarrow b\bar{b})_{\rm SM}$, i.e.\ the partial width
for $h \to b \bar b$ normalized to its SM value, where the other term
in \refeq{effcoupling} and further genuine loop corrections are taken 
into account (see text).
}
\label{fig:smallaleff}
\end{figure}
%%%%%%%%%%%%%%%%%%%%% FIGURE %%%%%%%%%%%%%%%%%%%%%%%%%%%%%%%%%%%%%%%%%%

The effects of the higher-order corrections to the couplings of 
the lightest Higgs boson to the down-type fermions appear very
pronounced in the ``small $\aeff$''~scenario~\cite{benchmark},
corresponding to the following choice of MSSM parameters:
\BEA
&& \mt \, = \, 174.3 \gev, \quad 
\msusy \, = \, 800 \gev, \quad
\Xt^{\rm FD}  = \, -1100 \gev \non \\
&& \mu \, = \, 2.5 \, \msusy, \quad
M_2 \, = \, 500 \gev, \quad
\Ab \, = \, \At, \quad
\mgl \, = \, 500 \gev~.
\label{smallaeff}
\EEA

In the upper row of \reffi{fig:smallaleff} we show the ratio 
$\sin^2 \aeff/\CQb$ in the $\MA$--$\tb$ plane, evaluated in the 
``small $\aeff$'' scenario. As explained above, replacing $\alpha$ by
$\aeff$ in the tree-level couplings of \refeq{treecouplings} 
gives rise to regions in the $\MA$--$\tb$ plane where the effective 
coupling of the lightest Higgs boson to the down-type fermions is
significantly suppressed with respect to the Standard Model. The region
of significant suppression of $\sin^2 \aeff/\CQb$ as evaluated with 
\fh1.0, i.e.\ without the inclusion of the sbottom corrections
beyond one-loop order, is shown in the upper left plot.
The upper right plot shows the corresponding result as evaluated with
\fh1.3, where the \order{\alb\als} and \order{\alb (\als\tb)^n}
corrections as well as the new \order{\alt^2} ones are included.
The new corrections are seen to have a drastic impact on the region
where $\sin^2 \aeff/\CQb$ is small. While without the new corrections a 
suppression of 70\% or more occurs only in a small area of
the $\MA$--$\tb$-plane for $20 \lsim \tb \lsim 40$ and $100 \gev \lsim
\MA \lsim 200 \gev$, the region where $\sin^2 \aeff/\CQb$ is very small
becomes much larger once these corrections are included. It now reaches
from $\tb \gsim
15$ to $\tb > 50$, and from $\MA \gsim 100 \gev$ to $\MA \lsim 350
\gev$. The main reason for the change is that the one-loop
\order{\alb} corrections to the Higgs-boson mass matrix, which for large
$\tb$ would prevent $\aeff$ from going to zero, are heavily suppressed
by the resummation of the \order{\alb (\als\tb)^n} corrections in the
bottom Yukawa coupling. This kind of suppression depends strongly on
the chosen MSSM parameters, and especially on the sign of $\mu$. 

In order to interpret the physical impact of the effective coupling
shown in the upper row of \reffi{fig:smallaleff}, the $\Delta_d$ terms
in \refeq{effcoupling} as well as further genuine loop corrections
occuring in the $h \to b \bar b$ process have to be taken into account.
The effect of these contributions can be seen from the plot in the lower
row of \reffi{fig:smallaleff}, where the ratio $\Gamma(h \rightarrow
b\bar{b})_{\rm MSSM} \,/\,\Gamma(h\rightarrow b\bar{b})_{\rm SM}$ is
shown, which has been evaluated by including all terms of
\refeq{effcoupling} as well as all further corrections described in 
\citere{hff}. The region where the partial width 
$\Gamma(h \rightarrow b\bar{b})$ within the MSSM is suppressed compared
to its SM value is seen to be somewhat reduced and 
shifted towards smaller values of $\MA$ as compared to the region where
$\sin^2 \aeff/\CQb$ is small.

%%%%%%%%%%%%%%%%%%%%%%%%%%%%%%%%%%%%%%%%%%%%%%%%%%%%%%%%%%%%%%
%%%%%%%%%%%%%%%%%%%%%%%%%%%%%%%%%%%%%%%%%%%%%%%%%%%%%%%%%%%%%%

\section{Estimate of the theoretical accuracy of the Higgs-boson mass
determination}
\label{sec:hocorr}

The prediction for $\mh$ in the MSSM is affected by two kinds of
uncertainties, parametric uncertainties from the experimental errors of
the input parameters and uncertainties from unknown higher-order
corrections. While currently the parametric uncertainties dominate over
those from unknown higher-order corrections, as the present
experimental error of the top-quark mass of about $\pm 5$~GeV induces an
uncertainty of $\De\mh \approx \pm 5$~GeV~\cite{tbexcl}, at the next
generation of colliders $\mt$ will be measured with a much higher
precision, reaching the level of about 0.1~GeV at an $e^+e^-$
LC~\cite{teslatdr}. Thus, a precise measurement of $\mh$ will provide a
high sensitivity to SUSY loop effects and will in this way allow
a stringent test of the MSSM, provided that the uncertainties from 
unknown higher-order corrections are sufficiently well under control.

Given  our present knowledge of the two-loop contributions to the
Higgs-boson self-energies, the theoretical accuracy reached in the 
prediction for the $\cp$-even Higgs-boson masses is quite advanced. 
However, obtaining a complete two-loop result for the 
Higgs-boson masses and 
mixing angle requires additional contributions that are not yet available. 
In this section we discuss the possible effect of the missing two-loop 
corrections, and we estimate the size of the higher-order (i.e.\ three-loop) 
contributions.

%%%%%%%%%%%%%%%%%%%%%%%%%%%%%%%%%%%%%%%%%%%%%%%%%%%%%%%%%%%%%%
%%%%%%%%%%%%%%%%%%%%%%%%%%%%%%%%%%%%%%%%%%%%%%%%%%%%%%%%%%%%%%

\subsection{Missing two-loop corrections}
\label{sec:misstwol}

It is customary to separate the corrections to the Higgs boson
self-energies into two parts: i) the momentum-independent part,
namely the contributions to the self-energies evaluated at zero
external momenta, which can also be computed in the effective
potential approach; ii) the momentum-dependent corrections, i.e.\ the
effects induced by the dependence on the external momenta of the
self-energies, that are required to determine the poles of the
$h,H$-propagator matrix.  

All the presently known two-loop contributions are computed at zero
external momentum, and moreover they are obtained in the so-called
gaugeless limit, namely by switching off the electroweak gauge
interactions. The two approximations are in fact related, since the
leading Yukawa corrections are obtained by neglecting both the momentum 
dependence and the gauge interactions. In order to systematically
improve the result beyond the approximation of the leading Yukawa terms,
both effects from the gauge interactions and the momentum dependence
should be taken into account.

In the limit where the momentum dependence and the gauge interactions
are neglected, the only missing contributions are the
mixed two-loop Yukawa terms, \order{\alt \alb}, and the
\order{\alb^2} corrections (and analogous contributions proportional to
the Yukawa couplings of the other fermions and sfermions, which however
are expected to give significantly smaller contributions than the third
generation quarks and scalar quarks). As in the case of the \order{\alb \als}
corrections, they can be numerically relevant only for large values of
$\tb$ and of the $\mu$ parameter. Concerning the \order{\alt\alb}
corrections, although their computation is not yet available, it is
plausible to assume that the most relevant contributions are connected
to the threshold effects in the bottom mass coming from chargino--stop
loops \cite{deltamb}, which are actually already implemented in \fh\
as an additional option. An estimate of the \order{\alb^2} corrections
for a simplified choice of the MSSM parameters is presented in
\citere{mhiggsEP4}. According to that discussion, their effect can be
at most comparable to that of the genuine \order{\alb \als}
corrections, the latter being exemplified in Fig.~\ref{fig:sbottom} by
the difference between the dashed and solid curves.

To try to estimate, although in a very rough way,
the importance of the various contributions we look at their 
relative size in the one-loop part. There, in the effective
potential part, the effect of the
\order{\alt} corrections typically amounts to an increase in $\mh$ of
40--60 GeV, depending on the choice of the MSSM parameters, whereas
the corrections due to the electroweak (D-term) Higgs--squark
interactions usually decrease $\mh$ by less than 5 GeV
\cite{mhiggsf1lA}.  Instead, the purely electroweak gauge
corrections to $\mh$, namely those coming from Higgs, gauge boson and
chargino or neutralino loops \cite{mhiggsf1lC}, are typically quite
small at one-loop, and can reach at most 5 GeV in specific regions of
the parameter space (namely for large values of $\mu$ and $M_2$). 
Concerning the effects induced by the dependence on the external
momentum, as a general rule we expect them to be more
relevant in the determination of the heaviest eigenvalue $m_H$ of
the Higgs-boson mass matrix, and when $\MA$ is larger than $\MZ$.
Indeed, only in this case the self-energies are evaluated at external
momenta comparable to or larger than the masses circulating into the
dominant loops. In addition, if $\MA$ is much larger than $\MZ$, the
relative importance of these corrections decreases, since the
tree-level value of $m_H$ grows with $\MA$.  In fact,
the effect of the one-loop momentum-dependent corrections
on $\mh$ amounts generally to less than 2~GeV.

Assuming that the relative size of the two-loop contributions follows
a pattern similar to the one-loop part, we estimate that the two-loop
diagrams involving D-term interactions should induce a variation in
$\mh$ of at most 1--2 GeV, while we expect those with pure gauge
electroweak interactions to contribute to $\mh$ not very
significantly, probably of the order of 1 GeV or less. Given the
smallness of the one-loop contribution it seems quite unlikely that
the effect of the momentum-dependent part of the \order{\alt \als}
corrections to $\mh$, which should be the largest among this type of
two-loop contributions, could be larger than 1 GeV.  As already said,
the situation can, in principle, be different for the heavier
Higgs-boson mass. However, the corrections to $\mH$ are relatively small in
general and at \onel\ level for most parts of the MSSM parameter space
the momentum-dependent corrections are not particularly relevant. The
momentum-dependent corrections turn out to be more relevant in
processes where the $H$~boson appears as an external particle, see
\citere{eennH}.

Another way of estimating the uncertainties of the kind discussed above
is to investigate the renormalization scale dependence introduced via
the \drbar\ definition of $\tb$, $\mu$, and the Higgs field
renormalization constants. Varying the scale parameter between $0.5 \mt$
and $2 \mt$ gives rise to a shift in $\mh$ of about 
$\pm 1.5$~GeV~\cite{mhiggsrenorm}, in accordance with the estimates
above.

%%%%%%%%%%%%%%%%%%%%%%%%%%%%%%%%%%%%%%%%%%%%%%%%%%%%%%%%%%%%%%
%%%%%%%%%%%%%%%%%%%%%%%%%%%%%%%%%%%%%%%%%%%%%%%%%%%%%%%%%%%%%%

\subsection{Estimate of the uncertainties from unknown three-loop
corrections}

Even in the case that a complete two-loop computation of the MSSM
Higgs masses is achieved, non-negligible uncertainties will remain,
due to the effect of higher-order corrections. Although a three-loop
computation of the Higgs masses is not available so far, it is
possible to give at least a rough estimate for the size of these 
unknown contributions.

A first estimate can be obtained by varying the renormalization scheme
in which the parameters entering the two-loop part of the corrections
are expressed. In fact, the resulting difference in the numerical
results amounts formally to a three-loop effect. Since the
\order{\alt \als + \alt^2} corrections are particularly sensitive to
the value of the top mass, we compare the predictions for $\mh$
obtained using in the two-loop corrections either the top pole mass,
$m_t^{\rm pole} = 174.3 \gev$, or the SM running top mass
$\overline{m}_t$, expressed in the \msbar\ renormalization scheme,
i.e.
\BE \mtms \;\;\equiv \;\;\mt(\mt)^{\msbarm}_{\SM}
\;\;\simeq \;\; \frac{\mt^{\rm pole}}{1+ 4\,\als(\mt)/3 \pi
-\alt(\mt)/2\pi} ~.
\label{mtrun}
\EE
%
%%%%%%%%%%%%%%%%%%%%% FIGURE %%%%%%%%%%%%%%%%%%%%%%%%%%%%%%%%%%%%%%%%%%
\begin{figure}[t]
\begin{center}
\epsfig{figure=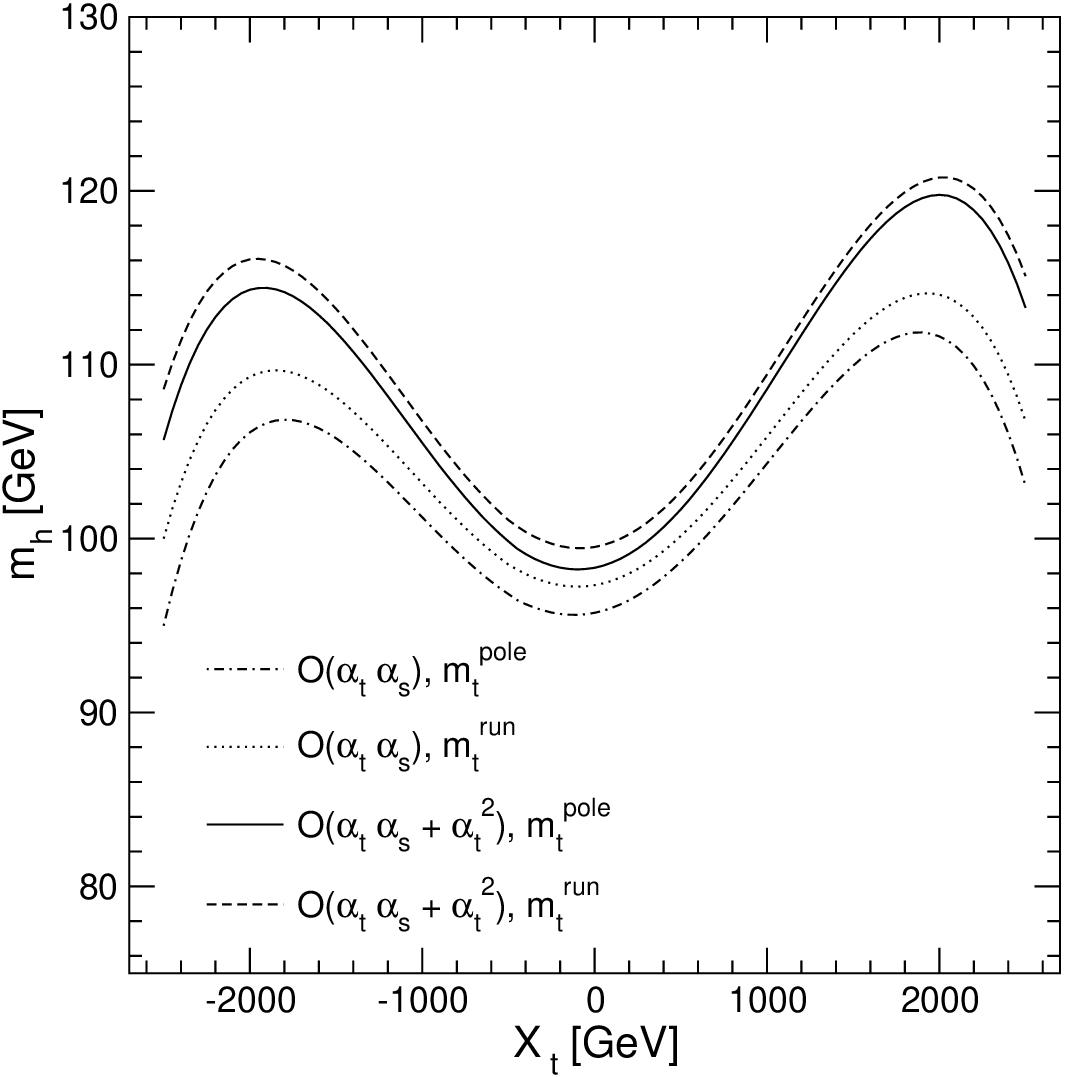,width=12cm}
\end{center}
\caption[]{$\mh$ as a function of $\Xt$, using either $\mt^{\rm pole}$
or $\overline{m}_t$ in the two-loop corrections.  The relevant MSSM
parameters are chosen as $\tb=3\,,\, \msusy = \MA = \mu = 1  \tev$
and $\mgl = 800 \gev$. For the different lines see text.} 
\label{fig:mtpolevsms}
\end{figure}
%%%%%%%%%%%%%%%%%%%%% FIGURE %%%%%%%%%%%%%%%%%%%%%%%%%%%%%%%%%%%%%%%%%%

Inserting appropriate values for the SM running couplings $\als$ and
$\alt$ we find $\overline{m}_t = 168.6 \gev$.  In
Fig.~\ref{fig:mtpolevsms} we show the effect of changing the
renormalization scheme for $\mt$ in the two-loop part of the
corrections. The relevant MSSM parameters are chosen as in
Fig.~\ref{fig:RGmeth}, i.e.~$ \tb = 3\,,\;\msusy = \MA = \mu = 1 \tev$
and $\mgl = 800 \gev$. The dot--dashed and dotted curves show the
\order{\alt \als} predictions for $\mh$ obtained using $\mt^{\rm
pole}$ or $\overline{m}_t$, respectively, in the two-loop
corrections.  The solid and dashed curves, instead, show the
corresponding \order{\alt \als + \alt^2} predictions for $\mh$.  The
difference in the two latter curves induced by the shift in $\mt$,
which should give an indication of the size of the unknown three-loop
corrections, is of the order of 1--1.5 GeV. However, as can be seen
from the figure, the effect of the shift in $\mt$ partially cancels
between the \order{\alt \als} and \order{\alt^2} corrections and there
is no guarantee that such a compensating effect will appear again in the
three-loop corrections.

An alternative way of estimating the typical size of the leading
three-loop corrections makes use of the renormalization group
approach.  If all the supersymmetric particles (including the $\cp$-odd
Higgs boson $A$) have mass $\msusy$, and $\beta = \pi/2$, the effective
theory at scales below $\msusy$ is just the SM, with the role of the
Higgs doublet played by the doublet that gives mass to the up-type
quarks. In this simplified case, it is easy to  
apply the techniques of \citeres{mhiggsRG1,mhiggsRG2} in order to obtain
the leading logarithmic corrections to $m_h$ up to three loops (see also
\citere{ahoang}). 
Considering for further simplification the case of zero stop mixing, 
we find:
\BE
\left(\Delta \mh^2\right)^{\rm LL} = 
\frac{3\,\alt\,\overline{m}_t^2}{\pi} \,t\,
\left[ \,1 + \left(\frac{3}{8}\,\alt - 2\,\als\right)\,\frac{t}{\pi}
+\left(\frac{23}{6}\,\als^2 - \frac{5}{4}\,\als\alt - 
\frac{33}{64}\,\alt^2 \right) \,\frac{t^2}{\pi^2}\, + ... \,\right]\;,
\EE
where $t = \log\,(\msusy^2/\overline{m}_t^2)$, $\overline{m}_t$ is
defined in \refeq{mtrun}, $\alt$ and $\als$ have to be
interpreted as SM running quantities computed at the scale $Q =
\overline{m}_t$, and the ellipses stand for higher loop contributions.
It can be checked that, for $\msusy = 1 \tev$, the effect of the
three-loop leading logarithmic terms amounts to an increase in $\mh$ of
the order of 1--1.5 GeV. If $\msusy$ is pushed to larger values, the
relative importance of the higher-order logarithmic corrections
obviously increases. In that case, it becomes necessary to resum the
logarithmic corrections to all orders, by solving the
appropriate renormalization group equations numerically.
Since it is unlikely that a complete three-loop
diagrammatic computation of the MSSM Higgs-boson masses will be
available in the near future, it will probably be necessary to combine
different approaches
(e.g.\ diagrammatic, effective potential and renormalization group), in
order to improve the accuracy of the theoretical predictions up to the
level required to compare with the experimental results expected at
the next generation of colliders.

\smallskip
To summarize this discussion, the uncertainty in the prediction
for the lightest $\cp$-even Higgs boson arising from not yet
calculatated three-loop and even
higher-order corrections can conservatively be estimated 
to be 1--2 GeV. From the various missing \twol\ corrections an
uncertainty of less than 3 GeV is expected.
However, it is extremely unlikely that all these effects would
coherently sum up, with no partial compensation among them. 
Therefore we believe that a realistic estimate of the uncertainty from
unknown higher-order corrections in the theoretical prediction for
the lightest Higgs boson mass should not exceed $3 \gev$.

%%%%%%%%%%%%%%%%%%%%%%%%%%%%%%%%%%%%%%%%%%%%%%%%%%%%%%%%%%%%%%
%%%%%%%%%%%%%%%%%%%%%%%%%%%%%%%%%%%%%%%%%%%%%%%%%%%%%%%%%%%%%%

\section{Conclusions}
\label{sec:concl}
In this paper we have discussed the phenomenological impact of recently
obtained results in the MSSM Higgs sector. The new corrections have now
been implemented in the code \fh, which in this way provides 
the currently most precise evaluation of the masses and mixing angles of
the $\cp$-even MSSM Higgs sector. We have analyzed the effects of
these new contributions by comparing the results obtained
with \fh1.3, the newest version of the code that incorporates all these
effects, with those derived with previous versions in which the
two-loop \order{\alt^2} corrections were implemented via a
renormalization group method 
and the sbottom corrections beyond one-loop were not taken into
account.  

As a result of these improvements the lower limits on $\Tb$
that can be derived by combining the theoretical upper bound on $\mh$
with its experimental lower bound, are significantly weakened. We have
also shown that, if the top-quark mass is increased by one standard
deviation, the constraint on $\Tb$ completely disappears if the other
MSSM parameters are such that $\mh$ takes its maximum value as a
function of $\tb$.  

Concerning the
coupling of the lightest Higgs boson to the down-type fermions, that is
related to the angle that diagonalizes the Higgs boson mass matrix
including higher order corrections, we have shown that the area in the
$\MA$--$\tb$-plane where its value gets suppressed by $70\%$ or more with
respect to the SM one is significantly modified.  

Finally, we have
given an estimate of the uncertainties related to the various kinds of 
two- and three-loop contributions to the Higgs boson masses that are
still unknown.  Since it is extremely unlikely that all these effects
would coherently sum up, the uncertainty in the theoretical prediction
for the lightest MSSM Higgs boson mass from unknown two- and three-loop
corrections should not exceed $3 \gev$.

%%%%%%%%%%%%%%%%%%%%%%%%%%%%%%%%%%%%%%%%%%%%%%%%%%%%%%%%%%%%%%%%%%
%
\section*{Acknowledgments}
G.D.\ and G.W.\ thank the Max-Planck-Institut f\"ur Physik in Munich
for its kind hospitality during part of this project. We furthermore 
thank M.~Frank for helpful discussions. This
work was partially supported by the European Community's
Human Potential Programme under contract 
HPRN-CT-2000-00149 (Collider Physics).
%
%%%%%%%%%%%%%%%%%%%%%%%%%%%%%%%%%%%%%%%%%%%%%%%%%%%%%%%%%%%%%%%%%%
%%%%%%%%%%%%%%%%%%%%%%%%%%%%%%%%%%%%%%%%%%%%%%%%%%%%%%%%%%%%%%%%%%
%
\subsection*{Note added:}
While finalizing this paper, \citere{spmartin} appeared, containing the 
two-loop result for $\mh$ obtained from the full two-loop effective 
potential of the MSSM~\cite{effpotfull}. The contributions included in 
\citere{spmartin} that go beyond the ones discussed in this paper,
namely momentum-independent contributions for non-vanishing gauge
interactions and D-term Higgs--squark interactions, turn out to yield a
shift in $\mh$ of 1~GeV or less in the numerical results given in 
\citere{spmartin}, thus confirming our estimate discussed in 
\refse{sec:misstwol}.
%
%%%%%%%%%%%%%%%%%%%%%%%%%%%%%%%%%%%%%%%%%%%%%%%%%%%%%%%%%%%%%%%%%%

%%%%%%%%%%%%%%%%%%%%%%%%%%%%%%%%%%%%%%%%%%%%%%%%%%%%%%%%%%%%%%
%%%%%%%%%%%%%%%%%%%%%%%%%%%%%%%%%%%%%%%%%%%%%%%%%%%%%%%%%%%%%%

\end{document}